\documentclass[aps,prx,twocolumn,showpacs,superscriptaddress,groupedaddress]{revtex4-2}

\usepackage{graphicx}% Include figure files
\usepackage{dcolumn}% Align table columns on decimal point
\usepackage{bm}% bold math
\usepackage{color} 
\usepackage{amsmath}
\newcommand{\pp}{\mathcal{\partial}}
\newcommand{\OO}{\mathcal{O}}

\newcommand{\pd}[2]{\frac{\partial #1}{\partial #2}}

\newcommand{\average}[1]{\left< #1 \right>}

\begin{document}

\preprint{APS/123-QED}

\title{Phase transition in non-Markovian animal exploration model with preferential returns}% Force line breaks with \\
% \thanks{A footnote to the article title}%

\author{Ohad Vilk$^{a,b,c}$,  Daniel Campos$^{d}$, Vicenç Méndez$^{d}$, Emmanuel Lourie$^{b,c}$, Ran Nathan$^{b,c}$, Michael Assaf$^{a,e,*}$}
 \affiliation{$^a$ Racah Institute of Physics, The Hebrew University of Jerusalem, Jerusalem, Israel,} 
 \affiliation{$^b$ Movement Ecology Lab, Department of Ecology, Evolution and Behavior, Alexander Silberman Institute of Life Sciences, Faculty of Science, The Hebrew University of Jerusalem, Jerusalem, Israel,} 
 \affiliation{$^c$ Minerva Center for Movement Ecology, The Hebrew University of Jerusalem, Jerusalem, Israel,} 
 \affiliation{$^d$ Grup de Física Estadística, Dept. de Física, Universitat Autónoma de Barcelona, 08193 Bellaterra (Barcelona), Spain,}
 \affiliation{$^{e}$Institute for Physics and Astronomy, University of Potsdam, Potsdam 14476, Germany}
\email{To whom correspondence should be addressed:  michael.assaf@mail.huji.ac.il}

%\date{\today}% It is always \today, today,
             %  but any date may be explicitly specified

\begin{abstract}
We study a non-Markovian and nonstationary model of animal mobility incorporating both exploration and memory in the form of preferential returns. We derive exact results for the probability of visiting a given number of sites and develop a practical WKB approximation to treat the nonstationary problem. We further show that this model adequately describes empirical movement data of Egyptian fruit bats (\textit{Rousettus aegyptiacus}) when accounting for inter-individual variation in the population. Finally, we study the probability of visiting any site a given number of times and derive the corresponding mean-field equation. Here, we find a remarkable phase transition occurring at preferential returns which scale linearly with past visits. Following empirical evidence, we suggest that this phase transition reflects a trade-off between extensive and intensive foraging modes.
\end{abstract}

%\keywords{Suggested keywords}%Use showkeys class option if keyword
                              %display desired
\maketitle

%\tableofcontents

\textit{Introduction.} 
Movement is a vital part of life and is key in a wide range of physical, biological and ecological systems. Theoretical and empirical frameworks are thus amply used to study the mechanisms underlying movement patterns in all organisms \citep{nathan2008movement}. In particular, individual-based modeling of movement has played a crucial role in studying dynamic systems across multiple spatiotemporal scales \citep{grimm2005individual, metzler2014anomalous, lourie2021memory}. These models can be applied to infer behaviors and draw causal links between observed phenomena and their underlying mechanisms beyond phenomenological description of the observed patterns \cite{gurarie2016animal}. 

Most theoretical models rely on Markovian assumptions to capture the properties of animal trajectories. However, memory and similar cognitive mechanisms are key to understanding patterns observed in animal foraging  \cite{fagan2013spatial, Bra15}. A range of taxa, from insects to primates, have been shown to exhibit spatial memory, and many of them are known to repeatedly return to previously visited sites as a part of their regular foraging strategies, mammals being a paradigmatic example \cite{polansky2015elucidating, ranc2021experimental, goldshtein2020reinforcement, toledo2020cognitive, lourie2021memory}. Notably, memory patterns must be properly balanced by the organisms with some level of behavioral plasticity to enhance flexibility and exploration (see, \textit{e.g.}, \cite{KROCHMAL2021161}). For all these reasons, correctly incorporating memory within stochastic models is an important research line for improving both predictive and descriptive tools of movement \cite{Mo04,Pa08,fagan2013spatial, riotte2015memory}. Indeed, it has been shown that heuristic models of memory can be derived from microscopic consideration for limited cases \cite{hod2004phase, schutz2004elephants}. At the same time, new experimental methods are allowing to disentangle, even under field conditions, memory effects on movement from other cognitive/perception mechanisms \cite{ranc2021experimental}. As long as such experimental and theoretical advances are able to nourish each other, new levels of detail in our understanding of living organisms can be potentially reached.

Dealing with memory and similar cognitive mechanisms still represents a significant theoretical challenge. Stochastic models that allow the individual to return to its original position (resettings) have attracted much attention recently~\citep{Evans2011, berger2015recursive, Evans2020}, but these only incorporate memory in an elementary way. More complex self-avoiding random walks or preferential returns (PR), where the individual returns to any previous location with a probability proportional to the number of previous visits have also been studied~\citep{song2010modelling, Boyer2014}. These models are non-Markovian (and typically also non-stationary), requiring that the individual identifies and keeps record of its entire trajectory. While the propagator of these models~\citep{Boyer2014b}, and some properties of relocation times~\citep{Campos2019}, have been computed, characterizing the revisits complete statistics to each particular location remains an open problem.

Here we study a non-Markovian and non-stationary mechanistic model of animal mobility, explicitly incorporating both the tendency of an individual to return to previously visited locations (PR) and the tendency to explore new sites.
Versions of this model have been used to model the mobility of humans \cite{song2010modelling} and monkeys \cite{boyer2012non}, the latter suggesting that monkey movements are non-random due to the use of memory and visitation patterns driven by resource availability. We generalize the model by accounting for stochasticity, incorporating inter-specific variability in the population, and allowing for nonlinear PR \cite{krapivsky2000connectivity}. We provide analytical solutions to this non-Markovian, non-stationary model that go well beyond previous \textit{mean-field} results. In particular, we present several approaches to analytically find the (non-stationary) probability of having visited $n$ sites at time $t$ and study the statistics of how revisits are distributed through the available locations. Our approach, based on the WKB (Wentzel–Kramers–Brillouin) approximation, is thus useful to deal with explicitly time-dependent and non-stationary problems. Remarkably, by allowing for nonlinear PR we find a phase transition as a function of the strength of the PR, where above some threshold the most visited site dominates the dynamics, receiving practically all new visits. We suggest that this phase transition reflects a balance between the tendency to return to known sites and the will to explore new ones \cite{berger2015recursive}. We further verify our predictions using simulations, and show that our theoretical results adequately describe the space use patterns and the revisitation dynamics of Egyptian fruit bats (\textit{Rousettus aegyptiacus}) to fruit trees. 

\textit{Model}. 
Our model consists of two elements~\cite{song2010modelling}: (i) exploration -- with probability $P_{new}$ the animal visits a new site, and (ii) PR -- 
with probability $1 - P_{new}$ the animal visits a previously visited site $i$ with probability $\Pi_i(m_i)$, where $m_i$ is the number of previous visits to site $i$. Following empirical evidence in humans and animals \cite{song2010modelling, boyer2012non} we assume that 
\begin{equation} \label{pnew}
P_{new} = q n^{-\beta}\;\;,\;\;\;\; \Pi_i(m_i)  =  \frac{m_i^\alpha}{\sum_{j =1}^{n}m_j^\alpha}.  
\end{equation}
Here $n$ is the number of previously visited sites, and $\beta > 0$ and $0< q <1$ control the animal's tendency to visit new sites indicating a power-law decay controlled by \textit{conformity exponent} $\beta$.
On the other hand, the \textit{PR exponent} $\alpha>0$, governs the tendency to return to a previously visited location. %Importantly, previous studies have only considered the case of $\alpha =1$.  % Thus, the model parameters are $q, \beta$ and $\alpha$. provide here some simple examples: e.g., for \beta = \alpha = 0 we recover a simple random walk. For \beta = -1, \alpha = 0 this is an exploding popuation in population dynamics. 
Furthermore, and without loss of generality, we order the sites by rank such that $i = 1$ is the most visited site with $m_1$ visits. 
Notably, we assume that the number of available sites is always larger than the number of visited sites, and that no significant resource depletion within a site occurs.

\textit{Cumulative number of sites}. 
The probability $P(n, t)$ of having visited $n$ sites in $t \gg 1$ time steps follows
\begin{equation}\label{master_eq_sites}
\partial P(n, t)/\partial t = q(n-1)^{-\beta} P(n-1, t) - q n^{-\beta} P(n, t). 
\end{equation}
Although this \textit{master equation} is interpreted here in the context of movement between spatially distributed sites, it can equivalently describe a birth-death process of population of size $n$, where the growth rate is proportional to $n^{-\beta}$
\footnote{Similar growth rates appear in self-inhibitory gene regulatory networks, where a protein inhibits its own growth, see \textit{e.g.}, Refs.~\cite{dublanche2006noise,roberts2015dynamics}}. 
In particular, for $\beta = 0$ the birth-death process is $\emptyset \xrightarrow{q} A$ and for $\beta = -1$ the birth-death process is $A \xrightarrow{q}  2A$. While these special cases have known exact solutions, in this manuscript we are primarily interested in the regime $\beta > 0$, which describes a growth which \textit{decreases} [or saturates, see Eq.~\eqref{pnew}] with the number of sites (or with the population size). To the best of our knowledge this regime has not been analytically studied.

An equation for the first moment $\left< n\right> = \sum_n n P(n, t)$ can be obtained from Eq.~\eqref{master_eq_sites} by multiplying the latter by $n$, summing over all $n$’s, and using the definition for $\average{n}$, resulting in 
$
    d\average{n}/dt = q \average{n^{-\beta}} 
$
which under the \textit{mean-field approximation}   $\average{n^{-\beta}} \simeq \average{n}^{-\beta}$ is solved by \cite{song2010modelling}
\begin{equation} \label{num_of_sites}
    \average{n} = [(1 + \beta) q t]^{1/(1+\beta) },
\end{equation}
predicting a power-law dependence on the time of measurement. A similar derivation for the second moment yields $\average{n^2} = \average{n}^2 + \average{n}$ such that the variance follows $\sigma_n^2 \equiv \average{n^2} - \average{n}^2 = \average{n}$, \textit{i.e.},  the variance of the number of sites is equal to the mean as in a Poisson process. This result, however, turns out to be inaccurate as it involves various uncontrolled assumptions, and is not consistent with simulations, as elaborated below.

An exact solution to Eq.~\eqref{master_eq_sites} can be found by Laplace transforming the equation and solving the resulting recurrence equation  \footnote{In the SI, Sec. S1.B we alternatively develop a a generating function approach, and Eq.~\eqref{master_eq_sites} becomes a fractional integro-differential equation, which is solvable in specific cases.}. The exact solution for $\beta \neq 0$ has the form [see Supplementary Information (SI) Sec. S1.A]
\begin{equation}
    \label{exact_result}
P(n,t)=(-1)^{n-1}n^{\beta}\sum_{k=1}^{n}\frac{k^{-\beta}e^{-qt/k^{\beta}}}{\prod_{j=1,j\neq k}^{n}\left(\frac{j^{\beta}}{k^{\beta}}-1\right)}.
\end{equation}
For special values of $\beta$ this result simplifies to 
\begin{equation} \label{special_cases}
    P(n, t) = \begin{cases}
    \frac{1}{(n-1)!}\sum _{k=1}^n \binom{n}{k}(\!-\!1)^{n-k} k^{n-1}  e^{-\frac{q t}{k}} & \beta = 1 \\
    \frac{1}{n!} (q t)^{n}e^{-q t} & \beta = 0
    \\
    e^{-nqt}\left(e^{qt}-1\right)^{n-1} & \beta = -1. 
    \end{cases}
\end{equation}
Although Eq.~\eqref{exact_result} is an exact solution, it is given in form of a summation of large terms of alternating sign, which converges due to a precise balance between the terms. Thus, in practice this result is very slow to converge for $n\gg 1$ in any finite-size system (\textit{e.g.}, python, matlab, and mathematica) and may result in a significant lack of accuracy. Moreover, virtually any approximation made in calculating the individual terms may cause large errors for $n \gg 1$ \cite{assaf2006spectral}. To circumvent these issues, we develop a time-dependent WKB approximation.  

\begin{figure}[t!]
\centering
\hspace{-3.5mm}\includegraphics[width=1.04\linewidth]{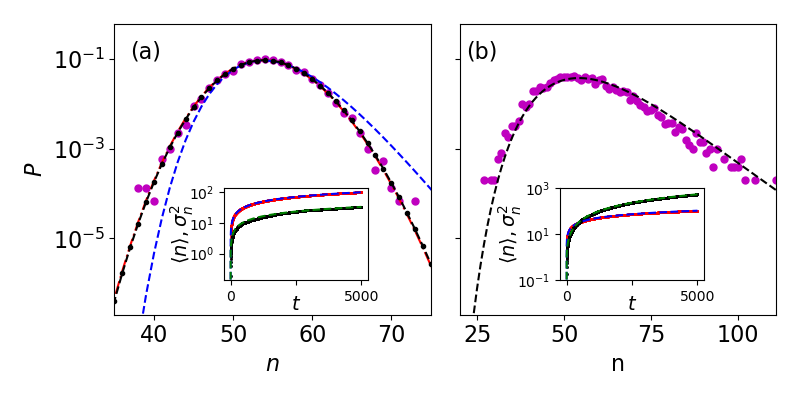}
 \vspace{-5mm}\caption{The probability $P(n, t)$ for $\beta = 1$ and $t = 1500$. (a) No variation in $\beta$ ($\sigma = 0$). We compare simulations (circles), exact result [black dashed-dotted line, Eq.~\eqref{exact_result}], WKB approximation [red dashed line, Eq.~\eqref{action_exact}], and WKB approximation at low energies [blue dashed line, Eq.~\eqref{ActionApprox}]. 
 (b) Variability in $\beta$ with $\sigma = 0.1$, compared to a numerical solution of Eq.~\eqref{P_I_def} (dashed line). Insets  show $\average{n}$ and $\sigma_n^2$ (red and black marks, respectively) versus $t$, compared with theory (dashed lines).
 } 
 \label{fig:examples}
\end{figure} 

% \subsection{\label{sec:TDWKB} Time dependent WKB}
\textit{Time dependent WKB}. 
In the limit of a large number of sites $n \gg 1$~\cite{dykman1994large, elgart2004rare,assaf2017wkb}, we substitute the time-dependent ansatz $P(n, t)\sim e^{- {\cal S}(n,t)}$ into Eq.~\eqref{master_eq_sites}. Neglecting terms of order $\OO(n^{-1})$ we obtain a classical Hamilton-Jacobi equation for the action function ${\cal S}(n, t)$:
% \begin{equation} \label{HamiltonJacobi equation}
   $ \pp_t {\cal S} = H(n, \pp_n {\cal S}) \equiv H(n, p) $
% \end{equation}
where we have defined the Hamiltonian 
$
    H(n, p) = q\left(1-e^{-p}\right) n^{-\beta }, 
$
and denoted $p = -\pp_n S$ as the conjugate momentum. Instead of directly solving the Hamilton-Jacobi equations, we use the Hamilton approach for the classical equations of motion 
\begin{eqnarray}
    &\dot{n} =  q e^{-p } n^{-\beta } \;,\;\;\;\dot{p} = \beta  q \left(1 - e^{-p }\right) n^{-\beta -1}. \label{dotn}
\end{eqnarray}
We write the action on a classical trajectory as ~\cite{elgart2004rare}: 
\begin{equation} \label{action_int}
    S = E t - \int_0^t p \dot{n} dt  = E t - \int^n p(n') dn'
\end{equation}
where the energy $E \equiv H[n(t), p(t)]$ is constant along a dynamical trajectory given by $p(n) = \log \left[q/(q-E n^{\beta})\right]$. To find the energy we solve the equation of motion \eqref{dotn} on a given dynamical trajectory on which the energy is constant. After some algebra (SI, Sec. S1.C) this yields 
\begin{eqnarray} \label{action_exact}
    P(n, t)&& \sim e^{- \average{n} \mathcal{S}(x)} \;,\;\;\; \mathcal{S}(x) = \frac{f(x) x^{-\beta }}{\beta +1}  \\ &&+x f(x)^{-1/\beta } B\left[f(x);1+1/\beta ,0\right] + x \log(1-f(x)),\nonumber
\end{eqnarray}
with $x \equiv n/\average{n}$ and $f(x) = 1-x^{\beta } (\beta  (x-1)+x)$.  Here, $B(z;a, b)=\int_0^z u^{a-1}(1-u)^{b-1}du$ is the incomplete beta function. This calculation of the probability of having visited $n\gg 1$ sites at time $t$, is one of our main results. In the low energy limit, $E\ll 1$, $S(x)$ becomes (SI)  
\begin{equation} \label{ActionApprox}
       \mathcal{S}(x) \simeq \left[(2 \beta +1) x^{-2 \beta -1} \left(x^{\beta +1}-1\right)^2\right]/\left[2 (\beta \!+\!1)^2\right]\!,
\end{equation}
which can be shown to solve the Hamilton-Jacobi equation in the limit $|x - 1| \ll 1$. In Fig.~\ref{fig:examples}a  we find good agreement between the exact result for the PDF [Eq.~\eqref{exact_result}],  time-dependent WKB approximation [Eqs.~(\ref{action_exact},\ref{ActionApprox})], and  simulations (see also Fig.~S1 in the SI). Here the exact result and WKB approximation are practically indistinguishable, whereas the low energy approximation predicts the PDF well only in its Gaussian vicinity. Notably, for $n\gg 1$ the accuracy of the exact result  rapidly deteriorates due to  summation of (alternating) very large terms and accumulation of errors, making the time-dependent WKB approach highly advantageous in this case~\cite{assaf2006spectral}.

Equation~\eqref{action_exact} predicts a different variance than that predicted by the mean-field approach above. The variance can be found by approximating $\mathcal{S}(x)$ in the Gaussian vicinity of $n = \average{n}$. Doing so, we find $\mathcal{S}(x) \simeq \left(\beta + 1/2\right) (x-1)^2$, which yields a variance of $\sigma_n^2 = \average{n}/|\mathcal{S}''(x)|_{x=1} = \average{n}/(1 + 2 \beta)$. This entails that the distribution is narrower by a factor of ($1 + 2 \beta$) than that predicted by the moment equations. 
In the inset of Fig.~\ref{fig:examples}a both the average [Eq.~\eqref{num_of_sites}] and the variance of number of sites show good agreement with simulations. 
 
To account for between-individual variation, we generalize our model by allowing  different $\beta$ values acoss individuals. Assuming $\beta$ is sampled from a normal distribution $\mathcal{N}(\beta_0, \sigma^2)$ with mean $\beta_0$ and variance $\sigma^2 \ll 1$ (indicating the inter-individual variability in $\beta$ around $\beta_0$, see  empirical results analysis below), $P_n(t)$ satisfies
\begin{equation} \label{P_I_def}
    P(n, t) = \frac{1}{\sqrt{2 \pi \sigma^2}}\int_{-\infty}^\infty P_\beta(n, t) e^{-\frac{(\beta - \beta_0)^2}{2 \sigma^2}}d\beta,
\end{equation}
where $P_\beta(n, t)$ is the probability for a given $\beta$, see \textit{e.g.}, Eq.~\eqref{action_exact}. Although analytical progress is possible in the limit of $\sigma\!\ll\! 1$~(SI, Sec. S1.D), Eq.~\eqref{P_I_def} can generally be solved numerically (Fig.~\ref{fig:examples}b). 
Notably, we checked that while variability in a population ($\sigma\! >\! 0$) will not significantly affect the mean number of sites, it dramatically affects the variability across a population~(SI, Fig. S2).

\vspace{4mm}
\textit{Statistics of number of visits to a site}. 
Having computed the statistics of number of sites, we now turn to study the probability $W_i(m_i, t)$ of having $m_i$ visits at time $t$ to an already visited site $i$, which follows 
\begin{eqnarray} \label{Wi_t}
\pd{W_i}{t} = (1-P_{new})  \left[   \Pi_i(m_i - 1) W_i(m_i-1, t) \right. \nonumber \\  \left. - \Pi_i(m_i)  W_i(m_i, t) \right], 
\end{eqnarray}
where $P_{new}$ and $\Pi_i$ are given by Eq.~(\ref{pnew}). In general, $\Pi_i$ depends on the number of visits to other sites, such that Eq.~\eqref{Wi_t} couples between different sites. Below, we focus on the  limit $t \gg 1$, where $P_{new}\!\to\! 0$ (SI, Sec. S2).

\begin{figure}[t!]
\centering
\hspace{-3.5mm}\includegraphics[width=1.04\linewidth]{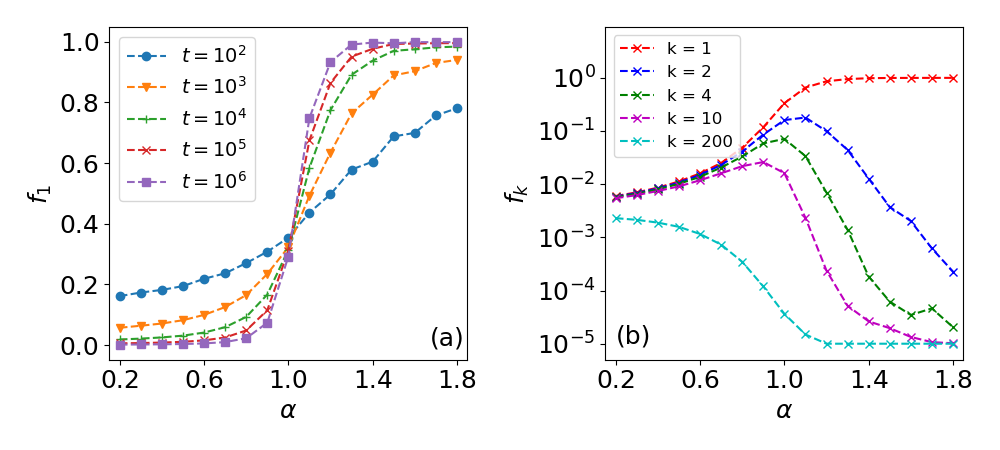}
\vspace{-5mm}
 \caption{(a) The average frequency of visits to the most visited site $f_1$ versus $\alpha$, for $\beta = 1$ (simulations). Each curve corresponds to a given number of visits $t$ (see legend).
 (b) $f_k$ for different sites (see legend) for $\beta = 1$ and $t = 10^5$. 
 } 
 \label{fig:freq_site_1}
\end{figure}

\textit{The case of $\alpha = 1$}. 
In the limit $t \gg 1$ we have $\sum_i m_i = t$, namely, the total number of visits to all sites equals the total number of time steps $t$. Thus, $\Pi_i(m_i) = m_i/t$ and $P_{new} = 0$~\footnote{Here, Eq.~\eqref{Wi_t} is similar to the master equation in Ref. \cite{Boyer2014}. However, they only consider the case of $\beta = 0$.}.
Equation~\eqref{Wi_t} is then solved by 
$
    W_i(m_i, t) = t_i t^{-m_i} \left(t-t_i\right){}^{m_i-1}, 
$
where $t_i$ is defined as the first time site $i$ is visited, $m_i(t = t_i) = 1$. 
The average is then $\average{m_i} = t/t_i$, while the variance $\sigma_{m_i}^2 = \average{m_i^2}  - \average{m_i}^2 = t \left(t-t_i\right)/t_i^2 \sim t^2$.

\textit{The case of $\alpha \neq 1$}. 
Here, we \textit{a priori} assume that        $\sum_{j=1}^{\average{n}} \average{m_j}^\alpha \sim t^\xi$, 
where $\xi$ is \textit{a priori} unknown and satisfies $\alpha<\xi<1$ (see SI, Sec. S2.A for proof). The average number of visits to any site $i$ can then be shown to asymptotically follow 
$\average{m_i} \sim t^{(1-\xi)/(1-\alpha)} [1 + \OO(t^{\xi-1})]$.  
Plugging this back into the sum over $\average{m_j}^\alpha$ we find that $\xi =\xi_0(1 + \epsilon)$, where $\xi_0 = (1+ \alpha\beta)/(1 + \beta)$ and $\epsilon \ll 1$ is an unknown function of $\alpha, \beta$ (Fig.~S3).  
For $1 - \alpha \ll \epsilon$ [$\beta = \OO(1)$], we further find $\average{m_i} \sim t^{\beta/(1+\beta)}$, which is independent of $\alpha$; yet, the condition $1 - \alpha \ll \epsilon$ breaks down as  $\alpha$ approaches 1 (SI, Sec. S2.B and Fig.~S4). Importantly, in the limit $t \gg 1$, for any $\alpha < 1$ we find that  all sites scale similarly with time. 
In contrast, for $\alpha > 1$ not all site scale similarly with time. Here we find $\average{m_i} \simeq t$ for $i = 1$, while $\average{m_i} \simeq C_i    [1 + \OO(t^{1-\alpha})]$ for $i > 1$, where $C_i = C_i(t_i)$ is a constant (SI, Sec. S2.C).

\begin{figure}[t!]
\centering
\includegraphics[width=1\linewidth]{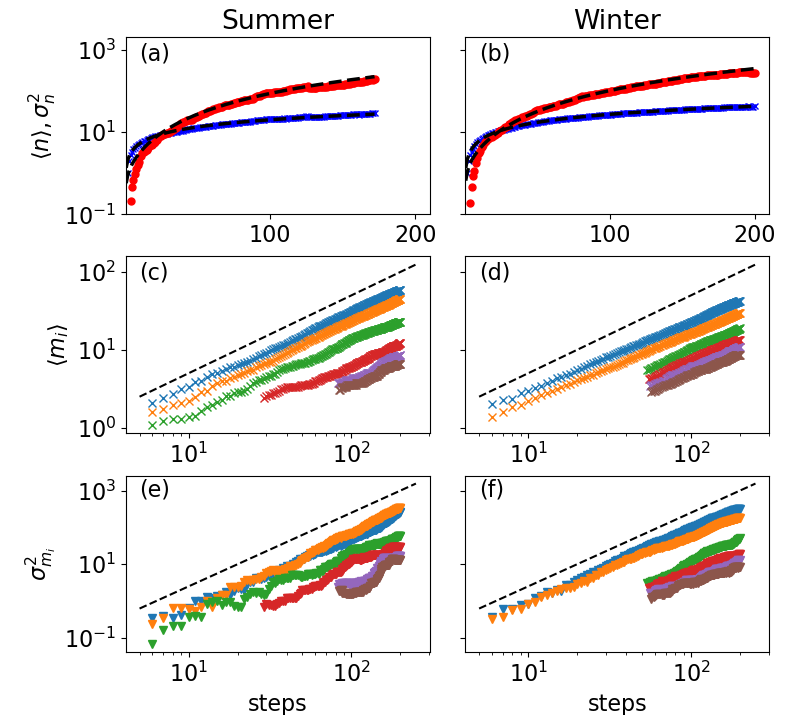}
 \caption{(a-b): The mean (blue marks) and variance (red marks) of number of sites visited by bats during summer and winter, compared to theoretical prediction (black dashed lines), with fitted values (a) $\beta_0 = 0.71, 0.53$ and (b) $\sigma = 0.21, 0.16$, respectively. These are averaged over an ensemble of 38 (a) and 53 (b) bats. (c-d): The mean number of visits $\average{m_i}$ to the six most visited sites, $i = 1,..., 6$ (different colors mark different sites), averaged over the same ensembles as in (a) and (b). Black dashed lines  $\average{m_i} \sim t$ correspond to the theoretical prediction for $\alpha = 1$. (e-f): Variance of number of visits $\sigma_{m_i}^2$ to the same six sites, averaged over the same ensembles for summer (e) and winter (f). Black dashed lines $\sigma_{m_i}^2 \sim t^2$ correspond to the theoretical prediction for $\alpha = 1$.  } 
 \label{fig:bats}
\end{figure}

These results reveal a phase transition at $\alpha = 1$ (see also SI, Sec. S2.D), where for weak PR ($\alpha < 1$) the frequency of visits to the most visited site $f_1 \equiv \average{m_1}/\sum_{j=1}^{\average{n}} \average{m_j}$ is only a small fraction of the total number of visits, while for strong PR ($\alpha > 1$) $f_1$ approaches 1 as $t$ is increased and site 1 dominates (Fig.~\ref{fig:freq_site_1}a). Importantly, in addition to the phase transition for $f_1$, the next most visited sites ($f_k$, for $k = 2, 3, ...$) peak around $\alpha = 1$ (Figs.~\ref{fig:freq_site_1}b and S5). Here, as $\alpha$ approaches 0 the number of visits to all sites becomes similar, while for $\alpha >1$ these visitation frequencies tend to zero. 

\textit{Movement of fruit bats}.
To study the relevance of our model for real-life systems and to obtain insights into the phase transition, we compare our predictions to resource use patterns and visitation dynamics of wild fruit bats tracked by ATLAS during winter and summer \footnote{To have proper statistics and ensure no significant resource depletion, we analyse 10-day periods for each bat. For details on the species, seasonality and the segmentation and fitting procedures, see SI Sec. S3 and \cite{lourie2021memory}. For details on ATLAS see \cite{weiser2016characterizing, toledo2020cognitive}}. In Fig.~\ref{fig:bats}a-b we fit the mean and variance of the number of visited sites (fruit trees) as a function of the number of movement steps (defined here as distinguishable movement between trees, see SI) to our theory \footnote{ Note that the power-law dependence for the mean has also been experimentally observed in humans \cite{song2010modelling} and monkeys \cite{boyer2012non}; however, these studies did not analyze fluctuations around $\langle n\rangle$.}. We find that during the summer $\beta_0$ and $\sigma$ are higher than during the winter, entailing lower rate of visits to new sites (higher levels of conformity) and larger inter-individual variability, respectively. 
Notably, many empirical studies have shown that individual preferences and decisions can affect movement and behavior, such that individuals do not have identical movement patterns \cite{dingemanse2002repeatability, dingemanse2013quantifying}. This may explain the inter-individual variability observed in both summer and winter.
In Fig.~\ref{fig:bats}c-f we show that during summer $\average{m_1} \sim t^{0.97}$ and $\average{m_2} \sim t^{0.99}$ (as well as $\sigma_{m_1}^2 \sim t^{1.94}$ and $\sigma_{m_2}^2 \sim t^{2.16}$), which matches the theory for $\alpha = 1$. In contrast, during winter $\average{m_1} \sim t^{0.89}$ and $\average{m_2} \sim t^{0.87}$ ($\sigma_{m_1}^2 \sim t^{1.96}$, $\sigma_{m_2}^2 \sim t^{1.80}$), which is consistent with $\alpha$ values slightly below 1. 
These seasonal differences may be attributed to the fact that bats during the summer feed of highly abundant and palatable fruits -- mulberries or figs with high levels of sugar content -- and hence do not need to explore for feeding alternatives (high $\beta_0$) and can strongly rely on a limited number of sites ($\alpha = 1$). Similarly, when food is abundant it makes sense that most bats will forage on the same bountiful locations, as they are not constrained by intense resource competition, thus reducing inter-individual variability in behaviour. In contrast, during winter there is less motivation to return to less favorable fruits (chinaberries) and a higher motivation to explore alternative trees, such as non-seasonal (unpredictable) fruit  from the Ficus family. 
 
In light of the phase transition at $\alpha = 1$, and in agreement with experimental results, we hypothesize that in animal movement the value of $\alpha$ will tend towards 1. This maximizes the frequencies of visits to preferred sites (Fig.~\ref{fig:freq_site_1}b), yet avoids an exclusive choice of a preferred site which occurs at $\alpha >1$ (see Fig.~\ref{fig:freq_site_1}a). In this manner the animal combines intensive search patterns (committing to few site) with extensive searches (returning to all sites with some probability), a balance which is essential to optimize between energy expenditure and risk management \cite{Bartumeus2014,Nolting2015, berger2015recursive}. Indeed, for fruit bats we find $\alpha \approx 1$, and similar results were obtained for humans \cite{song2010modelling} and monkeys \cite{boyer2012non}. Furthermore, the strategy of avoiding an exclusive site resembles bet-hedging strategies, \textit{e.g.}, bacterial persistence~\cite{Balaban2004}. In addition, the value of $\alpha$ may be correlated to the total number of known sites: for large $\beta_0$ (few sites, summer) the bats will show stronger PR, while for smaller $\beta_0$ (more sites, winter) the strategy may tend towards a more uniform visitation rate to all trees. 

More generally, we expect our results to also provide key insights into revisit dynamics in other areas like human mobility \cite{scbr21,doalca21,scdo21}, COVID-19 spread \cite{gaci21}, human migration \cite{chli20} and languages dynamics \cite{wa18}.

\section*{Acknowledgements}
OV and MA were supported by the Israel Science Foundation Grant No. 531/20. MA was also supported by the Humboldt Research Fellowship for Experienced Researchers of the Alexander von Humboldt Foundation. Fieldwork with ATLAS was funded by the Minerva Foundation, the Minerva Center for Movement Ecology, and grants ISF-1259/09, ISF-965/15 and GIF 1316/15 to RN. Finally, we thank Yoav Bartan, Anat Levi, David Shohami and other members of the Minerva Center for Movement Ecology for their valuable help with fieldwork. 

\bibliography{references}% Produces the bibliography via BibTeX.

\clearpage

\setcounter{equation}{0}
\setcounter{figure}{0}
\setcounter{table}{0}
\setcounter{page}{1}
\setcounter{section}{0}
\makeatletter
\renewcommand{\theequation}{S\arabic{equation}}
\renewcommand{\thefigure}{S\arabic{figure}}
\renewcommand{\thesection}{S\arabic{section}}

\onecolumngrid 

\LARGE 
\begin{center}
Supplementary Information: Phase transition in non-Markovian animal exploration model with preferential returns
\end{center}
\normalsize

Here we provide additional details and  results to support the derivations presented in the main text. 
Below, the notations and acronyms are the same as in the main text and the equations and figures refer to those therein.

\section{ Cumulative number of sites}
In this section we detail the exact solution and the WKB approximation to Eq.~(2) of the main text
\begin{equation}\label{master_eq_sitesSI}
    \frac{dP(n, t)}{dt} = q(n-1)^{-\beta} P(n-1, t) - q n^{-\beta} P(n, t). 
\end{equation} 

\subsection{\label{sec:laplace_solution} Solution by Laplace transform}
Here we derive Eqs.~(4) and (5) of the main text by Laplace transforming Eq.~\eqref{master_eq_sitesSI} and solving the resulting recurrence equation. First we define $J(n,t)=qP(n,t)/n^{\beta}$,
which turns Eq.~\eqref{master_eq_sitesSI} into:
\begin{equation}
\frac{n^{\beta}}{q}\frac{dJ(n,t)}{dt}=J(n-1,t)-J(n,t).\label{eq:J}
\end{equation}
Transforming (\ref{eq:J}) by Laplace in time and considering the
initial condition $P(n,t=0)=\delta_{n,1}$, where $\delta_{a,b}$
is the Kronecker delta, we obtain the recurrence
equation
\begin{equation}
\hat{J}(n,s)=\frac{1}{1+\frac{sn^{\beta}}{q}}\hat{J}(n-1,s)+\frac{1}{1+\frac{sn^{\beta}}{q}}\delta_{n,1},\label{eq:Jt}
\end{equation}
where $s$ is the Laplace variable and $\hat{J}(n,s)$ stands for
the Laplace transform of $J(n,t)$ defined as follows:
\[
\hat{J}(n,s)=\mathcal{L}_{s}\left[J(n,t)\right]=\int_{0}^{\infty}e^{-st}J(n,t)dt.
\]
Multiplying (\ref{eq:Jt}) by $\prod_{j=0}^{n}\left(1+\frac{sj^{\beta}}{q}\right)$
it has the form
\begin{equation}
A(n,s)=A(n-1,s)+\delta_{n,1}\prod_{j=0}^{n-1}\left(1+\frac{sj^{\beta}}{q}\right),\label{eq:A}
\end{equation}
where $A(n,s)=\hat{J}(n,s)\prod_{j=0}^{n}\left(1+\frac{sj^{\beta}}{q}\right)$
has been introduced. Since $P(n\leq0,t)=0$ one has $A(0,s)=0$ and
using (\ref{eq:A}), $A(n\geq1,s)=1$. Inserting this result into
(\ref{eq:Jt}) the solution to the master equation in the Laplace
domain reads
\begin{equation}
\hat{P}(n,s)=\frac{n^{\beta}}{q\prod_{j=0}^{n}\left(1+\frac{sj^{\beta}}{q}\right)}=\frac{q^{n-1}}{[(n-1)!]^{\beta}}\frac{1}{\prod_{j=1}^{n}\left(s+\frac{q}{j^{\beta}}\right)}.\label{eq:Pl}
\end{equation}
Now we need to invert (\ref{eq:Pl}) by Laplace. To do this we use
the Heaviside expansion theorem \cite{AS08}
\begin{equation}
\mathcal{L}_{t}^{-1}\left[\frac{1}{f(s)}\right]=\sum_{k=1}^{n}\frac{e^{-a_{k}t}}{f'(s=-a_{k})},\qquad f(s)=\prod_{j=1}^{n}(s+a_{j})\label{eq:ilt},
\end{equation}
where the prime symbol stands for the derivative with respect to $s$.
Making use of the property
\[
f'(s=a_{k})=\prod_{j=1,j\neq k}^{n}(a_{j}-a_{k}),
\]
and (\ref{eq:ilt}), one readily obtains for any $\beta \neq 0$
\begin{eqnarray}
\mathcal{L}_{t}^{-1}\left[\frac{1}{\prod_{j=1}^{n}\left(s+\frac{q}{j^{\beta}}\right)}\right]=\sum_{k=1}^{n}\frac{e^{-qt/k^{\beta}}}{\prod_{j=1,j\neq k}^{n}\left(\frac{q}{j^{\beta}}-\frac{q}{k^{\beta}}\right)}=(-1)^{n-1}(n!)^{\beta}q^{1-n}\sum_{k=1}^{n}\frac{k^{-\beta}e^{-qt/k^{\beta}}}{\prod_{j=1,j\neq k}^{n}\left(\frac{j^{\beta}}{k^{\beta}}-1\right)}.\label{eq:ilt2}
\end{eqnarray}
Finally, plugging (\ref{eq:ilt2}) into (\ref{eq:Pl}) the exact solution for $\beta\neq0$
has the form
\begin{equation}
    \label{exact_resultSI}
P(n,t)=(-1)^{n-1}n^{\beta}\sum_{k=1}^{n}\frac{k^{-\beta}e^{-qt/k^{\beta}}}{\prod_{j=1,j\neq k}^{n}\left(\frac{j^{\beta}}{k^{\beta}}-1\right)}, 
\end{equation}
which is Eq.~(4) of the main text, and is valid for any $\beta \neq 0$. For $\beta=0$, Eq.~\eqref{eq:Pl} reduces to
$$
\hat{P}(n,s)=\frac{1}{q\left(1+\frac{s}{q}\right)^{n+1}}, 
$$
which after inversion by Laplace coincides with Eq.~(5) of the main text
\begin{equation}\label{poiss}
   P(n, t) = \frac{ (q t)^{n}}{n!}e^{-q t}.
\end{equation}

\subsection{Solution by generating function}
In some special cases it is more convenient  to solve Eq.~\eqref{master_eq_sitesSI} using the generating function approach. Particularly, in the limit of a large number of sites, the equation in the generating function domain becomes a fractional integro-differential equation for $0< \beta \leq 1$, see below. 

We define the generating function $G(z, t) = \sum_{n} z^n P(n, t)$, such that 
\begin{equation} \label{PfromG}
    P(n, t) = \frac{1}{n!}\frac{\pp^n G(z, t)}{\pp z^n}|_{z =0},  
\end{equation}
with initial condition $P(n, 0) = \delta_{n, 1} \Rightarrow G(n,0) = z$. 
Substituting Eq.~\eqref{master_eq_sitesSI} into the definition of $G(n, t)$ yields
\begin{eqnarray} \label{GeneratingFunctionEq}
\pd{G}{t} &&= q \sum_{n} \left[z^n(n-1)^{-\beta} P(n-1, t) -  z^n n^{-\beta} P(n, t)\right] 
= q (z -1) \sum_{n} z^n n^{-\beta} P(n, t)  \nonumber\\
&& \simeq q (z -1) \sum_{n} D^{-\beta}_z z^{n - \beta} P(n, t) 
  =  q (z -1)  D^{-\beta}_z \left[z^{-\beta} G(z, t)\right],
\end{eqnarray}
where $D^{-\beta}_z$ is the Riemann-Liouville fractional integral defined by $_a D^{-\beta}_z f(z) = \frac{1}{\Gamma(\beta)}\int_a^z (z - \xi)^{\beta -1}f(\xi)d\xi$, and we used the following relations 
\begin{equation}
D^{-\beta}_z z^{n - \beta} = z^n  \frac{\Gamma(n-\beta+1)}{\Gamma(n+1)} = z^n n^{-\beta}[1 + \OO(1/n)] . 
\end{equation}
Here, the equality on the left is valid for $\beta \leq 1$ while the approximation on the right holds for $n \gg 1$ and is exact in the special cases of $\beta = 0, 1$. Notably, for $\beta < 0$, \textit{i.e.}, when the growth rate \textit{increases} in $n$, Eq.~\eqref{GeneratingFunctionEq} is a fractional differential equation for $G$, while for $\beta > 0$, \textit{i.e.}, when the growth rate \textit{decreases} in $n$, it is a fractional integro-differential equation. Although Eq.~\eqref{GeneratingFunctionEq} is hard to solve analytically for general $\beta$, and the direct method presented above is more suited, it can be solved for $\beta = -1, 0, 1$. %\red{consider adding further discussion on the fractional equation - why is it useful, and why is the derivative replaced by an integral for $\beta > 0$? }

\subsubsection{Solution for $\beta = 0$}
For $\beta =0$, Eq.~\eqref{GeneratingFunctionEq} simplifies to 
$
    \pp_t G(z,t) =  q (z -1)  G(z, t)
$
and is accordingly solved by 
$
    G(z, t) = z e^{-q t(1 - z)}
$. 
Using Eq.~\eqref{PfromG} we find that $P(n,t)$ follows a Poisson distribution~(\ref{poiss}). In particular, in this case we have $\average{n} = \sigma_n^2 = q t$.

\subsubsection{Solution for $\beta = -1$}
For $\beta = -1$ a similar derivation to Eq.~\eqref{GeneratingFunctionEq} yields a partial differential equation  for the generating function:
$
    \pp_t G(z,t) =  q (z -1)z \pp_z [ G(z, t)]
$, 
whose solution is
$G(z, t) = z/[z+ e^{q t}(1-z)]$~\cite{gardiner2009stochastic}.
Using Eq.~\eqref{PfromG} we find 
\begin{equation}
    P(n, t) = e^{-n q t} \left(e^{q t}-1\right)^{n-1}. 
\end{equation}
The average number of sites here is $\average{n} = e^{qt}$, \textit{i.e.}, we find exponential growth, as expected for a growth rate that is linear in $n$. Here, the variance is $\sigma_n^2 = e^{q t} \left(e^{q t}-1\right) \simeq e^{2qt} = \average{n}^2$, which is significantly broader than that of the Poisson distribution. This result also agrees with Eq.~(5) of the main text. 

\subsubsection{Solution for $\beta = 1$}
For $\beta = 1$, Eq.~\eqref{GeneratingFunctionEq} is rewritten in explicit integro-differential form: 
\begin{equation} \label{Geq_beta1}
\pd{G}{t} =  q (z -1)  \int_0^z y^{-1} G(y, t)dy. 
\end{equation}
Here, we Laplace transform Eq.~\eqref{Geq_beta1} in time
\begin{equation} 
u G(z, u) =  z +  q (z -1)  \int_0^z y^{-1} G(y, u)dy, 
\end{equation}
to obtain an integral equation. This equation can be solved iteratively by the Neumann series method \cite{arfken1999mathematical} to give:
\begin{equation} \label{GeneratingFunLaplace}
G(z, u) = \frac{1}{u} \left[(z-1) e^{\frac{q z}{u}} \left(\frac{q z}{u}\right)^{-\frac{q}{u}} \gamma \left(\frac{p+u}{u},\frac{q z}{u}\right)+z\right],
\end{equation}
where $\gamma(a, b)$ is the lower gamma function. 
Using Eq.~\eqref{PfromG} we can inverse Laplace transform Eq.~\eqref{GeneratingFunLaplace} to obtain 
\begin{equation} \label{exact_result_beta1}
    P(n, t) = \frac{1}{(n-1)!}\sum _{k=1}^n (-1)^{n-k} k^{n-1} \binom{n}{k} e^{-\frac{q t}{k}},
\end{equation}
in agreement with Eq.~(5) of the main text. 

\subsection{Time dependent WKB approximation}
Here we derive Eqs.~(8) and (9) of the main text. We employ the time-dependent %Wentzel–Kramers–Brillouin 
WKB approximation in the limit of a large number of sites $n \gg 1$~\cite{elgart2004rare,assaf2017wkb}. Substituting the time-dependent ansatz $P(n, t)\sim e^{- S(n,t)}$ into Eq.~\eqref{master_eq_sitesSI} 
and neglecting terms of order $\OO(n^{-1})$ we obtain a classical Hamilton-Jacobi equation for the action function $S(n, t)$:
\begin{equation} \label{HamiltonJacobi equation}
    \pd{S}{t} = H(n, \pd{S}{n}) \equiv H(n, p)\,,\;\;\;\;\;\;\;H(n, p) = q\left(1-e^{-p}\right) n^{-\beta },
\end{equation}
where $H$ is the Hamiltonian and  $p = -\pp_n S$ is the conjugate momentum. Instead of directly solving the Hamilton-Jacobi equations, we use the Hamilton approach for the classical equations of motion [Eq.~(6)]
\begin{eqnarray}
    \dot{n} =  q e^{-p } n^{-\beta } \label{dotnSI}\,,\;\;\;\;\;
    \dot{p} = \beta  q \left(1 - e^{-p }\right) n^{-\beta -1}. \label{dotrho}
\end{eqnarray}
We write the action on a classical trajectory as~\cite{elgart2004rare}: 
\begin{equation} \label{action_intSI}
    S = E t - \int_0^t p \dot{n} dt  = E t - \int^n p(n') dn'
\end{equation}
where the energy $E \equiv H[n(t), p(t)]$,
is constant along a dynamical trajectory given by $p(n) = \log \left[q/(q-E n^{\beta})\right]$.
To find the energy we solve the equation of motion \eqref{dotnSI} on this given dynamical trajectory, which yields
\begin{equation} \label{dotn2}
    \dot{n} = q n^{-\beta }-E.
\end{equation}
For $n\gg 1$ and $\beta > 0$ the right hand side of Eq.~\eqref{dotn2} varies very slowly with time [$\OO(n^{-\beta})$] (as shown below, the energy $E$ also scales as $n^{-\beta}$), such that the solution for Eq.~\eqref{dotn2} can be approximated as $n = ( q n^{-\beta }-E) t + C$.  Here, $C$ is a slowly-varying function of time, and includes constants such that the energy corresponding to the mean-field solution $n =\average{n}$ obeys $E(n = \average{n}) = 0$~\cite{elgart2004rare}. Having shown that $\average{n} \sim t^{1/(1+\beta)}$, we find $C = \average{n}\beta/(1+\beta)$, which  indeed varies with time slower than $t$. Substituting this back into the equation for $n$ and solving for the energy yields 
\begin{equation} \label{energy}
    E = q \average{n}^{-\beta} \left\{x^{-\beta }+[\beta -(\beta +1) x]\right\}
\end{equation}
where we have expressed $t$ in terms of $\average{n}$ and defined $x \equiv n/\average{n}$. 
Substituting the energy \eqref{energy} into Eq.~\eqref{action_intSI} and solving the integral yields 
\begin{eqnarray} \label{action_exactSI}
   S(n, t) = \average{n} \mathcal{S}(x)\quad, \quad\quad \mathcal{S}(x) = \frac{f(x) x^{-\beta }}{\beta +1}+x f(x)^{-1/\beta } B\left[f(x);1+\frac{1}{\beta },0\right]  + x \log(1-f(x))
\end{eqnarray}
where $B(z;a, b)$ is the incomplete beta function, defined as $B(z;a,b)=\int_0^z u^{a-1}(1-u)^{b-1}du$, and we define $f(x) = 1-x^{\beta } (\beta  (x-1)+x)$. This result coincides with Eq.~(8) of the main text and is valid in the limit of $n \gg 1$. 

\subsubsection{Low energy solution}
\begin{figure}[t!]
\centering
\includegraphics[width=0.65\linewidth]{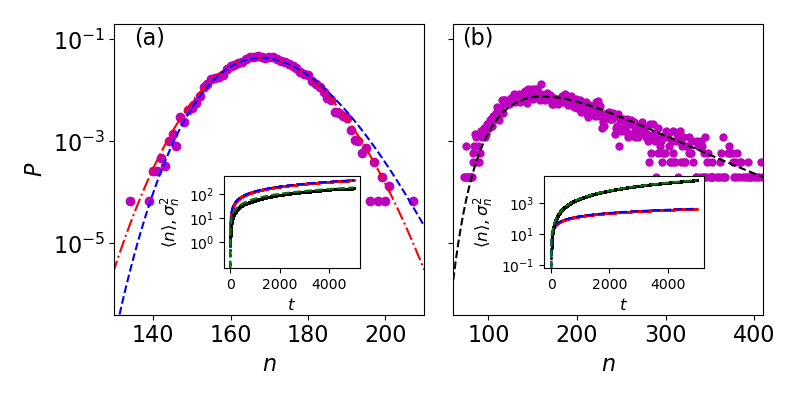}
 \caption{The probability $P(n, t)$ for $\beta = 0.5$ and $t = 1500$. (a) No variation in $\beta$ ($\sigma = 0$). We compare simulations (circles), WKB approximation [red dashed line, Eq.~\eqref{action_exact}], and WKB approximation at low energies [blue dashed line, Eq.~\eqref{ActionApprox}]. %In the inset are $\average{n}$ and $\sigma_n^2$ (red and black marks, respectively) as a function of $t$, both showing very good agreement with the theory (dashed lines). 
 (b) Variability in $\beta$ with $\sigma = 0.1$, compared to a numerical solution of Eq.~\eqref{P_I_def} (dashed line). In the insets of both panels are $\average{n}$ and $\sigma_n^2$ (red and black marks, respectively) as a function of $t$, showing very good agreement with the theory (dashed lines).} 
 \label{fig:example_beta05}
\end{figure}

To get better insight for the Gaussian vicinity of $P_n(t)$, we solve Eq.~\eqref{dotnSI} in the  low energy limit $E\ll 1$. This yields an approximated solution for the energy in the form 
\begin{equation} \label{energy_approx}
    E \simeq q \average{n}^{-\beta }  \frac{(2 \beta +1) x^{-2 \beta -1} \left(1 - x^{\beta +1}\right)}{\beta +1},
\end{equation}
where we have again expressed $t$ in terms of $\average{n}$. Equation~\eqref{energy_approx} is indeed small in the limit $|x - 1| \ll 1 $, which is the Gaussian vicinity of $P_n(t)$. By further approximating the dynamical trajectory as $p(n) \simeq E n^{\beta }/q + E^2 n^{2 \beta }/(2 q^2)$, we substitute this back into Eq.~\eqref{action_intSI}. Performing the integral and substituting Eq.~\eqref{energy_approx} yields the following action 
\begin{equation} \label{ActionApproxSI}
        \mathcal{S}(x) \simeq \frac{(2 \beta +1) x^{-2 \beta -1} \left(x^{\beta +1}-1\right)^2}{2 (\beta +1)^2},
\end{equation}
which coincides with Eq.~(9) of the main text. Equation~\eqref{ActionApprox} can be shown to solve Eq.~\eqref{HamiltonJacobi equation} in the limit $|x - 1| \ll 1$. 
In Fig.~1a and \ref{fig:example_beta05}a we compare the WKB solutions to simulations. Notably, while in Fig.~1a (main text) we are able to plot the exact results, in Fig.~\ref{fig:example_beta05}, due to the larger values of $n$, the exact result cannot be plotted with standard computational tools. %In Fig.~\ref{fig:numvisitsversusbeta} we further plot the mean number of sites visited at a given time, which decreases as a function of $\beta$. 

\subsection{Individual variability}
\begin{figure}[t!]
\centering
\includegraphics[width=0.45\linewidth]{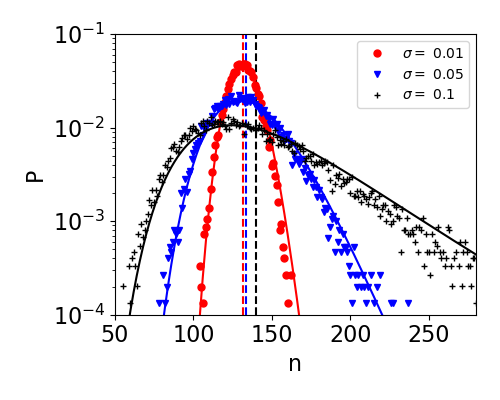}
 \caption{The probability distribution of number of sites visited at time $t = 1000$ for $\sigma = 0.01, 0.05, 0.1$ (see legend), based on 15000 simulations, compared to Eq.~\eqref{P_I_lowenergySI} (solid lines), where the integral is approximated numerically. Note that, the averages $\average{n}$ for each distribution, marked by vertical dashed lines, are only slightly affected by the change in $\sigma$.} 
 \label{fig:ind_var_dist_sim}
\end{figure}

Here we analytically solve Eq.~(10) of the main text in the limit of small variance $\sigma$. In the main text we write the probability of having visited $n$ sites at time $t$ as [Eq.~(10)] 
\begin{equation} \label{P_I_defSI}
    P(n, t) = \frac{1}{\sqrt{2 \pi \sigma^2}}\int_{-\infty}^\infty P_\beta(n, t) e^{-\frac{(\beta - \beta_0)^2}{2 \sigma^2}}d\beta,
\end{equation}
which can be numerically solved (Figs.~1b, \ref{fig:example_beta05}b and \ref{fig:ind_var_dist_sim}). Analytical progress can only be made in limit of small variance, $\sigma \ll 1/\sqrt{\average{n}_0}$, where $\average{n}_0 = [(1 + \beta_0) q t]^{1/(1+\beta_0)}$ is the mean number of sites given $\beta = \beta_0$.
For simplicity we focus on the small energy regime, yet similar calculations can be made with the full expression for the action [Eq.~(\ref{action_exact})]. Substituting Eq.~\eqref{ActionApprox} into Eq.~\eqref{P_I_defSI} yields 
\begin{equation}\label{P_I_lowenergySI}
    P(n, t) \sim \frac{1}{\sqrt{2 \pi \sigma^2}}\int_{-\infty}^\infty e^{-\frac{(\beta - \beta_0)^2}{2 \sigma^2} - \average{n}_0\mathcal{S}_{\beta}(n/\average{n}_0)}d\beta,
\end{equation}
where $\mathcal{S}_\beta(x)$ is given by Eq.~\eqref{ActionApprox}. This integral can be solved for $\sigma \ll 1/\sqrt{\average{n}_0}$, using the saddle point approximation, which yields 
\begin{equation}
    P(n, t) \sim e^{-\average{n}_0\mathcal{S}_{\beta_0}(n/\average{n}_0) +\average{n}_0^2 \sigma ^2 \mathcal{S}_1(n/\average{n}_0)},
\end{equation}
with 
\begin{eqnarray}
    S_1(x) =  \frac{x^{-4 \beta _0-2} \left(x^{\beta _0+1}-1\right)^2}{2 \left(\beta _0+1\right){}^6} \left[\beta _0 (1 + x^{\beta_0+1})- \left(\beta _0+1\right) \left(2 \beta _0+1\right) \ln (\average{n}_0 x) + 1\right]^2. 
\end{eqnarray}
Here, the mean number of sites obeys $\average{n} = \average{n}_0\left[1 + \OO(\sigma^2)\right]$, whereas the variance obeys  
\begin{eqnarray}
    \sigma_n = \average{n}_0\left\{\frac{1}{2 \beta _0+1}+ \average{n}_0 \sigma ^2\frac{ \left[\left(\beta _0+1\right) \ln (\average{n}_0)-1\right]^2}{\left(\beta _0+1\right){}^4}  + \OO(\average{n}_0^2 \sigma^4)\right\}.
\end{eqnarray}
Thus, while inter-individual variability will almost not affect the mean number of sites, it does significantly affect the variance of the number of sites (by a factor of $\average{n}_0$ compared to that of the mean), see Fig.~\ref{fig:ind_var_dist_sim}. 

\section{Statistics of number of visits to a site}
Here we provide details on the mean-field equation for the mean number of sites. In particular we explicitly derive and solve this equation for all $\alpha$ values in the limit of $t \gg 1$, and provide evidence of a phase transition at $\alpha = 1$. 
Our starting point is Eq.~(11) of the main text
\begin{eqnarray} \label{Wi_tSI}
\pd{W_i}{t} = (1-P_{new})  \left[   \Pi_i(m_i - 1) W_i(m_i-1, t) - \Pi_i(m_i)  W_i(m_i, t) \right], 
\end{eqnarray}
where $P_{new}$ and $\Pi_i$ are given by Eq.~(1) in the main text.

\subsection{The case of $\alpha = 1$}
In the main text we assumed that $P_{new} \to 0$ and solved Eq.~\eqref{Wi_tSI}. Here we provide a solution to the mean-field equation without neglecting $P_{new}$. 
In mean field, we obtain an equation for the first moment $\average{m_i}$ by multiplying Eq.~\eqref{Wi_tSI} by $m_i$ and summing over all $m_i$. For $\alpha = 1$ this yields
\begin{eqnarray} \label{rate_equation_full}
    \pd{\average{m_i}}{t} &&= (1 - P_{new}) \sum_{m_i = 1}^\infty \frac{m_i}{\sum_{j = 1}^{n}m_j} W_i(m_i, t) \simeq 
    \frac{\average{m_i}}{\sum_{j=1}^{\average{n}} \average{m_j}}(1- q\average{n}^{-\beta}), 
\end{eqnarray}
where we \textit{a priori} (to be checked \textit{a posteriori}) assume that $\sum_j m_j \gg m_i$ for any site $i$ such that the denominator can be taken out of the sum over $m_i$, and that $\sum_{j = 1}^{n}m_j \simeq \sum_{j = 1}^{n}\average{m_j}$. 
To find the value of $ \sum_{j=1}^{\average{n}} \average{m_j} \equiv Q$ we sum over both sides of Eq.~\eqref{rate_equation_full} to obtain a differential equation for $Q$: 
$
\pp_t Q =  (1- q\average{n}^{-\beta})
$, 
an equation which is solved by $Q = t - \average{n}$. 
Substituting this back into Eq.~\eqref{rate_equation_full} gives 
\begin{equation} \label{rate_equation_full2}
    \frac{d\average{m_i}}{dt} =  \frac{\average{m_i}}{t - \average{n}}(1- q\average{n}^{-\beta}),
\end{equation}
which is solved, assuming site $i$ is first visited at time $t_i$ [\textit{i.e.}, with an initial condition $\average{m_i}(t_i) = 1$], by \cite{song2010modelling}
\begin{equation} \label{m_i(t)}
    \average{m_i} = \frac{t - \average{n}}{t_i-\average{n}_{t_i}} \simeq \frac{t}{t_i}\,.
\end{equation}
Here $\average{n}_{t_i}$ is the average number of sites at time $t_i$, and on the right we approximated the solution for $t\gg 1$ and discarded terms of order $\OO (t^{1/(1+\beta)})$. This final result agrees with the one found in the main text. Importantly, as all sites have a linear dependence on $t$, we verify \textit{a posteriori} that $\sum_j m_j \gg m_i$ for any site $i$, as contribution from all visited sites will not diminish at long times.

\begin{figure}[t!]
\centering
\includegraphics[width=0.45\linewidth]{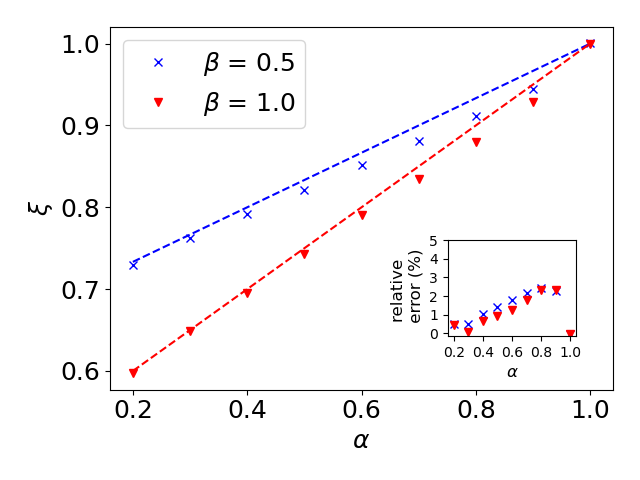}
 \caption{Comparison between the value of $\xi$ as obtained from 100 simulation of length $t = 10^5$ (symbols) to the theoretical prediction (dashed lines). Plotted for $\beta = 0.5$ (blue crosses) and $\beta = 1$ (red triangles). In the inset we plot the relative error between the predicted value and the one obtained in simulations.} 
 \label{fig:xi}
\end{figure} 

\subsection{The case of $\alpha < 1$}
For $\alpha < 1$ we solve the mean-field equation at $t \gg 1$ such that $P_{new} \to 0$, 
\begin{eqnarray} \label{RateEqNumVisits_smallalpha}
    \pd{\average{m_i}}{t} \simeq \sum_{m_i = 1}^\infty \frac{m_i^\alpha}{\sum_{j = 1}^{n}m_j^\alpha} W_i(m_i, t) \simeq \frac{\average{m_i}^\alpha}{\sum_{j = 1}^{n}\average{m_j}^\alpha} \simeq\frac{\left\langle m_{i}\right\rangle^\alpha}{A t^\xi}, 
\end{eqnarray}
where, similarly to the case of $\alpha = 1$, we \textit{a priori} assume that $\sum_j \average{m_j}^\alpha \gg \average{m_i}^\alpha$ for any site $i$, and we further assume $\sum_{j=1}^{\average{n}} \average{m_j}^\alpha = A t^\xi$ with $\alpha<\xi<1$ (to be proved \textit{a posteriori}, see below). Notably, such a scaling was found to hold in numerical simulations.
For initial condition $\average{m_i}(t = t_0) = 1$, Eq.~\eqref{RateEqNumVisits_smallalpha} is solved by 
\begin{equation} \label{m_i_general_alpha}
     \average{m_i} \simeq \left[1 + \frac{(\alpha -1) \left(t^{1-\xi }-t_i^{1-\xi }\right)}{A (\xi -1)}\right]^{1/(1-\alpha)}. 
\end{equation}
Note that the asymptotic scaling of this result at $t \gg t_i$ depends on the value of $\xi$, where for $\xi<1$, Eq.~\eqref{m_i_general_alpha} predicts an asymptotic scaling of $\average{m_i} \sim t^{(1-\xi)/(1-\alpha)} [1 + \OO(t^{\xi-1})]$, for all sites. Now, as all sites scale similarly with $t$, it is evident that $\sum_j \average{m_j}^\alpha \gg \average{m_i}^\alpha$, thus verifying our initial assumption. Using this solution for $\average{m_i}$, we find that up to some unknown factor $\sum_{j=1}^{\average{n}} \average{m_j}^\alpha \sim t^{1/(1+\beta)}t^{\alpha (1-\xi)/(1-\alpha)}$, entailing that $\xi = \alpha (1-\xi)/(1-\alpha) + 1/(1+\beta) = (1+ \alpha\beta)/(1 + \beta)$. 

In Fig.~\ref{fig:xi} we show that this prediction agrees with simulations for two different values of $\beta$, up to a maximum of $3\%$ relative error. However, we note that this relative error becomes crucial when $\xi$ is substituted back into the scaling $\average{m_i} \sim t^{(1-\xi)/(1-\alpha)}$ found above. Let us denote $\xi_0 = (1+ \alpha\beta)/(1 + \beta)$ such that $\xi = \xi_0(1-\epsilon)$, where $\epsilon \ll 1$ is a small correction that depends on $\alpha$ and $\beta$ (see Fig.~\ref{fig:xi}). Substituting $\xi$ into $\average{m_i} \sim t^{(1-\xi)/(1-\alpha)}$ one readily obtains $(1-\xi)/(1-\alpha) = \beta/(1+\beta) + \xi_0 \epsilon/(1-\alpha)$. Here, the approximation is valid only as long as $\beta/(1+\beta) \gg \epsilon/(1-\alpha)$, or alternatively $1 - \alpha \ll \epsilon$ [assuming $\beta = \OO(1)$]. As we numerically find that $\epsilon = \OO(10^{-1})$ this condition is hard to satisfy as $\alpha$ approaches 1, and in this regime it is preferable to find $\xi$ directly from simulations.  

\begin{figure}[t!]
\centering
\includegraphics[width=1\linewidth]{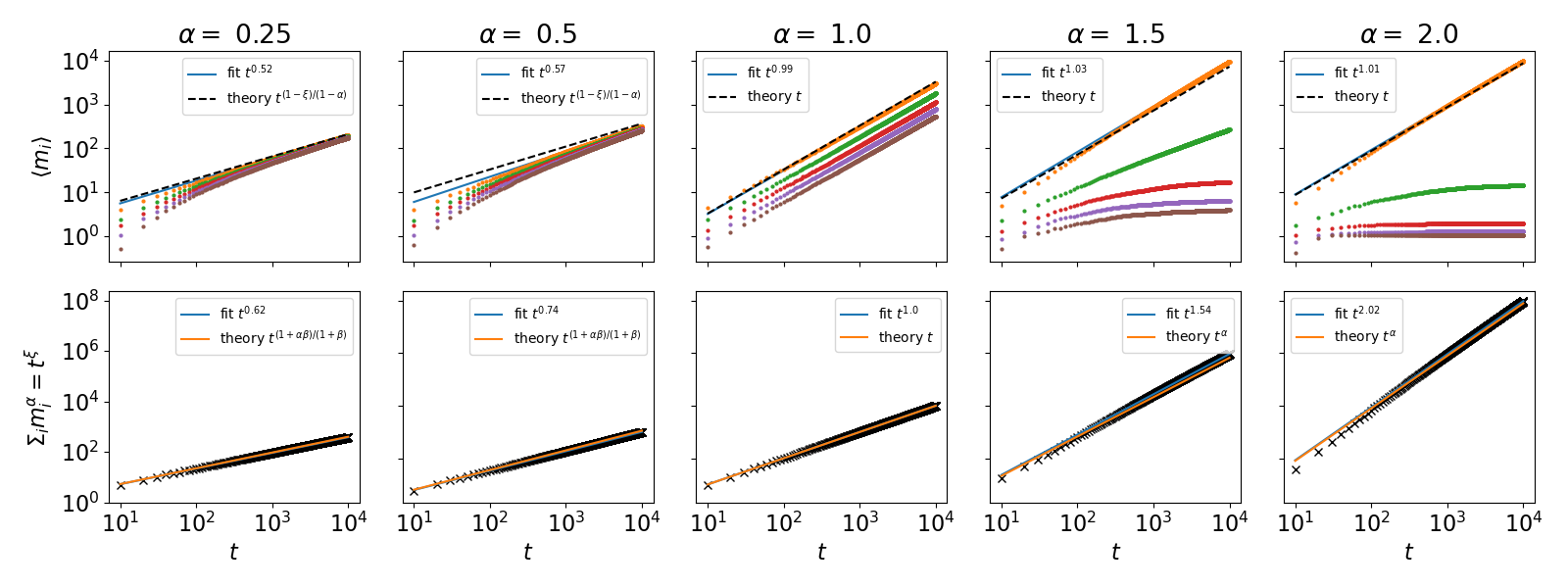}
 \caption{Upper panels: the mean number of visits to the five most visited sites for different values of $\alpha$ (different colors mark different sites). We fit the most visited site to a power law (blue solid lines, see legend) and plot the theoretical prediction (black dashed line, see legend).  Lower panels: measuring $\xi$ from simulations (black crosses), fit (blue solid line) and theory (orange solid line).} 
 \label{fig:mi_and_xi}
\end{figure}

In Fig.~\ref{fig:mi_and_xi} we plot the number of visits to the five most visited site for different values of $\alpha$. We show that for $\alpha < 1$ all sites converge to the same number of visits, while for $\alpha > 1$ the most visited site diverges from all other sites (see below). All $\alpha$ values show good agreement with the theory presented above. Similarly, in the bottom panels of \ref{fig:mi_and_xi}  we show evidence for our \textit{a priori} assumption that $\sum_{j=1}^{\average{n}} \average{m_j}^\alpha = A t^\xi$, again with good agreement to the theory.

\subsection{The case of $\alpha > 1$}
\begin{figure}[t!]
\centering
\includegraphics[width=0.65\linewidth]{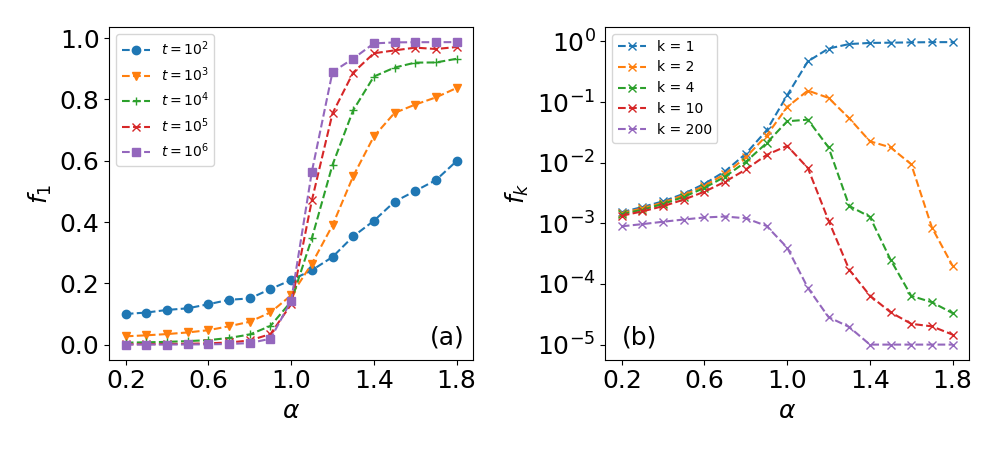}
 \caption{(a) The average frequency of visits to the most visited site $f_1$ versus $\alpha$, for $\beta = 0.5$ (simulations). Each curve corresponds to a given number of visits $t$ (see legend).
 (b) $f_k$ for different sites (see legend) for $\beta = 0.5$ and $t = 10^5$. 
 } 
 \label{fig:freq_site}
\end{figure}

Here, in contrast to the previous cases, in the limit of $t \gg 1$ we \textit{a priori} assume that $\average{m}_1^\alpha \gg \sum_{j=2}^{\average{n}} \average{m_j}^\alpha$, \textit{i.e.} at long times the most visited site dominates and contributions from all other sites diminish. We again obtain an equation for the first moment $\average{m_i}$ by multiplying Eq.~\eqref{Wi_tSI} by $m_i$ and summing over all $m_i$:
\begin{eqnarray} \label{RateEqNumVisits}
    \pd{\average{m_i}}{t} \simeq \sum_{m_i = 1}^\infty \frac{m_i^\alpha}{\sum_{j = 1}^{n}m_j^\alpha} W_i(m_i, t) \simeq 
    \begin{cases}
    1 
    %- \OO\left(\frac{\sum_{j=2}^{n}m_j^\alpha}{m_1^\alpha}\right), 
     & i = 1 \\
    \frac{\average{m_i}^\alpha}{\sum_{j = 1}^{n}\average{m_j}^\alpha}
    & i > 1, 
    \end{cases}  \label{mvmi}
\end{eqnarray}
where  we have separated the most visited site $i = 1$ from all other sites, in accord with the above assumption. For $i = 1$, Eq.~\eqref{RateEqNumVisits} with initial conditions $m_1(0) = 1$ is solved by $\average{m_1} \simeq 1 + t \simeq t$, \textit{i.e.} we predict a linear scaling with time. For all other sites we assume that $\sum_{j=1}^{\average{n}} \average{m_j}^\alpha \simeq \average{m_1}^\alpha \sim t^\alpha$. Plugging this into Eq.~\eqref{RateEqNumVisits} yields 
\begin{equation}
    \average{m_i} \simeq \begin{cases} 
    t & i = 1 \\
    \mbox{const}%\left(\frac{A'+t^{1-\alpha }-t_0^{1-\alpha }}{A'}\right)^{\frac{1}{1-\alpha }} 
    [1 + \OO(t^{1-\alpha})]& i > 1, 
    \end{cases}
\end{equation}
where $\mbox{const} \sim (1 - t_i^{1-\alpha})^{1/(1-\alpha)}$. Importantly, for $\alpha >1 $ and $\beta >0$ it follows that $\alpha > 1/(1+\beta)$, such that $ \average{m_1}^\alpha \sim t^\alpha \gg t^{1/(1+\beta)} \sim \sum_{j=2}^{\average{n}} \average{m_j}^\alpha$, thus verifying our initial assumption. 
As discussed in the main text, these results suggest a phase transition at $\alpha =1$, see Fig.~3 and \ref{fig:freq_site}, and the next subsection.

\subsection{Evidence of a phase transition}
Here we prove that there is a phase transition at $\alpha = 1$, with no \textit{a priori} assumptions on the solution (see previous sections). To this end, we define $Q \equiv \sum_{i=1}^{\average{n}} m_i^{\alpha}$, so the probability that the most visited site will be visited again in the next time step $t$ will be $p_{1}(t)=m_{1}^{\alpha}/Q $, and for any site in general we have $p_{i}(t)=m_{i}^{\alpha}/Q $.
Next, we write an expression for the expected value of $p_1(t)$ in the next time step:
\begin{equation}
  \average{p_{1}(t+1)} = \frac{m_{1}^{\alpha}+f_{\alpha}(m_{1})}{Q+f_{\alpha}(m_{1})} p_{1}(t) +  \sum_{i=2}^{\average{n}} \frac{m_{1}^{\alpha}}{Q+f_{\alpha}(m_{i})} p_{i}(t),
\label{average1}
\end{equation}
where we have defined $f_{\alpha}(m_{i}) \equiv (m_i +1)^{\alpha} - m_i^{\alpha}$. The first term on the right hand side represents the case for which the most visited site is visited in the next time step $t+1$, and the second term corresponds to the case where any other site is chosen instead.

In the case $\alpha =1$ we have $f_{1}(m_{i})=1$ for any $m_i$. Taking into account that $p_{1}(t)=m_{1}^{\alpha}/Q$, Eq.~(\ref{average1}) leads after some algebra to $\average{p_{1}(t+1)}= p_{1}(t)$ independently of the specific set of values $\{ m_i \}$ we have. Thus, the probability of revisiting the most visited site will be kept constant through time (and the same can be proved for any other site). 

For $\alpha > 1$ we note that  $f_{\alpha}(m_{i})$ increases monotonically with $m_{i}$. This, together with the fact that $p_{1}(t) = 1- \sum_{i=2}^{\average{n}} p_{i}(t)$ allow us to write the inequality
\begin{equation}
  \average{p_{1}(t+1)} > \frac{m_{1}^{\alpha}+f_{\alpha}(m_{1})}{Q+f_{\alpha}(m_{1})} p_{1}(t) +  \frac{m_{1}^{\alpha}}{Q+f_{\alpha}(m_{2})} (1- p_{1}(t) ).
\label{average2}
\end{equation}
Finally, introducing $p_{1}(t)=m_{1}^{\alpha}/Q $ into the previous inequality, after some algebra we obtain
\begin{equation}
  \average{p_{1}(t+1)} > \left[ 1+ \frac{(f_{\alpha}(m_{1})-f_{\alpha}(m_{2}) (Q-m_1^{\alpha})}{(Q+f_{\alpha}(m_{1}))(Q+f_{\alpha}(m_{2}))} \right] p_{1}(t).
\label{average3}
\end{equation}
We thus conclude that $\average{p_{1}(t+1)} > p_{1}(t)$ regardless of the specific set $\{ m_i \}$ we have. The probability of revisiting the most visited site thus always increases with time on average, leading eventually to its dominance  over the others.

For $\alpha <1$ we proceed in a similar manner. Here, $f_{\alpha}(m_{i})$ will decrease monotonically with $m_{i}$, so we can write
\begin{equation}
  \average{p_{1}(t+1)} < \left[ 1 - \frac{(f_{\alpha}(m_{2})-f_{\alpha}(m_{1}) (Q-m_1^{\alpha})}{(Q+f_{\alpha}(m_{1}))(Q+f_{\alpha}(m_{2}))} \right] p_{1}(t).
\end{equation}
This leads to $\average{p_{1}(t+1)} < p_{1}(t)$, such that for $\alpha<1$, on average the probability of revisiting the most visited site will decrease with time, thus leading to a much more homogeneous distribution of revisits among all sites available.

\section{Data collection and analysis}
The Egyptian fruit bat (\textit{Rousettus aegyptiacus}, EFB) is a long-lived, widely distributed Old World fruit bat \cite{kwiecinski1999rousettus}. Like other fruit bats, individual EFBs tend to feed on a small subset of available trees and repeatedly revisit them for weeks and even months \cite{egert2018resource, toledo2020cognitive} affirms that EFBs rely heavily on individual memory. Additionally, it has recently been shown that EFBs obtain a "cognitive map," which encompasses information about a large number of tree locations, suggesting that memory expands beyond the trees used at a given time \cite{toledo2020cognitive}. 
Bats were tracked at a 0.125Hz sampling rate for an average tracking period of 23.7 nights and up to 131 nights. The data also includes nearly all fruit trees in the study area (14,314 trees and 18,111 orchard trees), which enabled us to identify specific tree visits.

To segment the data into movements and tree visits, we first filtered raw EFB tracks for localization errors based on the covariance matrices attributed to each ATLAS fix \cite{gupte2020guide}. Localization that exceeded the highest realistic speed threshold for this species (20$\frac{m}{s}$) were removed. Visits to trees were defined based on track segmentation using the first-passage algorithm to determine the center of a "cloud of fixes" where the animal has spent a specified number of observations within a certain radius (for source code and details see https://github.com/ATLAS-HUJI/R/tree/master/ AdpFixedPoint). Finally, the median coordinates of each cloud were related to the closest tree in the dataset. 

To make the seasonal classification most relevant for bats' foraging, we defined winter and summer based on the known peak of fruiting periods of the main seasonal tree species the bats frequently visit in the study area. These are the mulberry (Morus negra) and common fig (Ficus carica) species during May-September (summer) and Chinaberry (Melia azedarach) during November-February (winter).     
During each fruiting period we used 10 day periods for each bat to ensure sufficient statistics (many of the bats did not have longer tracks during a single season). Based on the field work it is reasonable to assume that during such 10 day periods no significant resource depletion occurs. 
To fit between the data and the theory we used a standard least square fit procedure from python scipy package.

\end{document}